# Hot pressing to enhance the transport $J_c$ of $Sr_{0.6}K_{0.4}Fe_2As_2$ superconducting tapes

He Lin[1], Chao Yao[1], Xianping Zhang[1], Chiheng Dong[1], Haitao Zhang[1], Dongliang Wang[1], Qianjun Zhang[1], Yanwei Ma[1]*, Satoshi Awaji[2], Kazuo Watanabe[2], Huanfang Tian[3], and Jianqi Li[3]

[1]Key Laboratory of Applied Superconductivity, Institute of Electrical Engineering, Chinese Academy of Sciences, PO Box 2703, Beijing 100190, China

[2]High Field Laboratory for Superconducting Materials, Institute for Materials Research, Tohoku University, Sendai 980-8577, Japan

[3]Beijing National Laboratory for Condensed Matter Physics, Institute of Physics, Chinese Academy of Sciences, Beijing 100190, China

(*) Author to whom any correspondence should be addressed. E-mail: ywma@mail.iee.ac.cn

**Abstract**

High-performance $Sr_{0.6}K_{0.4}Fe_2As_2$ (Sr-122) tapes have been successfully fabricated using hot pressing (HP) process. The effect of HP temperatures (850-925 °C) on the c-axis texture, resistivity, Vickers micro-hardness, microstructure and critical current properties has been systematically studied. Taking advantage of high degree of c-axis texture, well grain connectivity and large concentration of strong-pinning defects, we are able to obtain an excellent $J_c$ of $1.2 \times 10^5$ A/cm$^2$ at 4.2 K and 10 T for Sr-122 tapes. More importantly, the field dependence of $J_c$ turns out to be very weak, such that in 14 T the $J_c$ still remains ~ $1.0 \times 10^5$ A/cm$^2$. These $J_c$ values are the highest ever reported so far for iron-pnictide wires and tapes, achieving the level desired for practical applications. Our results clearly strengthen the position of iron-pnictide conductors as a competitor to the conventional and $MgB_2$ superconductors for high field applications.

The relatively high critical temperature $T_c$, ultrahigh upper critical fields $H_{c2}$ and low anisotropy $\gamma$ of iron-pnictides make them potential competitors to conventional Nb-based and $MgB_2$ superconductors. Particularly, single crystals and thin films of 122-type iron-pnictides ($AeFe_2As_2$, Ae = alkali or alkali earth elements) maintained high $J_c$ over 1 MA/cm$^2$ up to 10 T at 4.2 K and $H_{c2}$ greater than 100 T[1-5], allowing the possibility of constructing superconducting magnets, especially for nuclear magnetic resonance (NMR) magnets and for accelerators. Since large-scale high field applications require long length of conductors, the simple powder-in-tube (PIT) method has been rapidly developed to fabricate $Ba_{1-x}K_xFe_2As_2$ (Ba-122) and $Sr_{1-x}K_xFe_2As_2$ (Sr-122) wires and tapes[6-8]. The critical current density in superconductors is a central topic of research. However, the primary obstacle to practical applications is the low global $J_c$ value in wires and tapes because the polycrystalline superconductors suffer from the disadvantages of defects, impurity phase and high-angle grain boundaries (GBs).

In order to enhance the connectivity between grains, *ex-situ* PIT technique was employed to improve homogeneity and reduce pores. However, the mass density of superconducting cores inside flat-rolled tapes is still low[9, 10]. It was found that the mechanical deformation such as hot pressing (HP) and cold pressing (CP) could significantly increases the transport $J_c$ by improving core density and grain alignment[11-14]. Importantly, hot pressing could not only considerably densify the superconducting core but also effectively prevent the formation of cracks in comparison with CP. Recently, the transport $J_c$ value up to 0.1 MA/cm$^2$ at 4.2 K and 10 T has been obtained in Sr-122 tapes by simple HP method[12]. However, the correlations among HP processing conditions, superconducting properties and microstructure are still not systematically studied. Therefore, understanding how to further enhance $J_c$ in Sr-122 tapes by optimizing HP process is quite important. In this work, we prepared the hot-pressed $Sr_{1-x}K_xFe_2As_2$/Ag tapes by optimizing the HP temperature. A transport $J_c$ of $1.2 \times 10^5$ A/cm$^2$ at 10 T and 4.2 K is achieved, which is higher than those have been reported elsewhere in any iron-pnictide wires and tapes. The reason of $J_c$ enhancement is also analyzed by means of structural and transport measurements in combination with scanning electron microscopy (SEM) observations. We also measured the distribution of local grain orientation and misorientation angles by electron backscatter diffraction (EBSD) technique. The flux pinning and GBs are analyzed by transmission electron microscopy (TEM) method to clarify how the high transport $J_c$ is achieved in iron-based superconductors.

**Results**

The bulk XRD patterns for the hot-pressed Sr-122 samples with different HP temperatures are shown in Fig. 1. For comparison, the data for randomly orientated powder is also included. It can be seen that all tape samples exhibit a well developed Sr-122 phase, with a small amount of AgSrAs phase. On the other hand, compared with randomly orientated powder, the relative intensity of the (00$l$) peaks with respect to that of the (103) peak is strongly enhanced for our HP samples. The strong c-axis orientation of Sr-122 grains is achieved after hot pressing. Fig. 2 exhibits the temperature dependence of the resistivity for HP tapes. Similar high onset temperature $T_c$ ~36.2 K and small transition width ~0.7 K are observed for all Sr-122 samples. In addition, the electrical resistivity curves display the characteristic temperature dependence of 122-type pnictides with a high ratio of room-temperature and normal-sate resistivity[15-17]. The resistivity measurement demonstrates good quality of our HP tapes, which encourages a further detailed investigation on transport $J_c$-$B$ properties and microstructure.

Fig. 3 presents the magnetic-field dependence of transport $J_c$ at 4.2 K for the Sr-122 tapes hot-pressed at different temperatures. The applied fields up to 14 T were parallel to the tape surface. As is evident from the figure, the $J_c$ increases monotonically with the increase of HP temperatures up to 900 °C, and then rapidly decreases when further increasing temperature to 925 °C. The best Sr-122 tapes exhibit a large $J_c$ of $1.2 \times 10^5$ A/cm$^2$ at 10 T and 4.2 K, which is the highest value ever reported so far for iron-pnictide wires and tapes. On the other hand, it has been reported that 122-type iron-pnictides have strong intrinsic pinning potential[4]. Here, we observe that transport $J_c$ exhibits a power-law dependence $J_c \propto B^{-\alpha}$ on the magnetic field. The field dependences of $J_c$ are similar for all HP tapes, and we get a α value of

~0.10±0.02 at 4.2 K and high fields. The value is much smaller than those observed in $Nb_3Sn$ and $MgB_2$ conductors[18-20]. This means that the $J_c$ of 122-type pnictide conductors has a very weak dependence in high fields at liquid helium temperatures. Actually, even in high field up to 14 T, the $J_c$ of HP900 tapes remains about $1.0 \times 10^5$ A/cm$^2$.

The insight into the effects of HP temperature on superconducting properties can be obtained from the HP-temperature dependence of $J_c$, c-axis texture parameter $F$, residual resistivity ratio $RRR$ and Vickers hardness $Hv$ values, as shown in Figs. 4(a-d). Firstly, the c-axis texture parameter $F$ in Fig. 4(b) can be quantified from XRD data by the Lotgering method with $F = (\rho-\rho_0)/(1-\rho_0)$, where $\rho = \sum I(00l)/I(hkl)$, $\rho_0 = \sum I_0(00l)/I_0(hkl)$[21]. $I$ and $I_0$ are the intensities of each reflection peak ($hkl$) of XRD patterns for the textured and randomly oriented samples, respectively. The $F$ values are about 0.52, 0.53, 0.58 and 0.57 for HP850, HP875, HP900 and HP925 tapes, respectively. Clearly, the $F$ value increases with the increase of HP temperature until ≥900°C. It is noted that the planar Sr-122 grains are prone to rotate along the tape surface by the external pressure force. At higher sintering temperature (≥ 900 °C) with external pressure, a drastic re-crystallization reaction may easily induce better grain alignment[22]. Secondly, the residual resistivity ratio $RRR = \rho(300K)/\rho(40K)$ for hot-pressed samples with different HP temperature is summarized in Fig. 4(c). The $RRR$ values of tapes prepared at 850, 875, and 900 °C are monotonic as a function of the HP temperature, which increase when increasing HP temperature. The better re-crystallization reaction, which introduces better grain connectivity, is believed to be responsible for this[19, 20]. In contrast, there is a decrease of the $RRR$ for HP925 tapes, which may be ascribed to more large structure defects[23]. Thirdly, the core density is an important factor affecting the transport $J_c$ of iron-pnictide wires and tapes. Researchers usually use Vickers hardness as an indication of the core density[11, 13]. Therefore, Vickers micro-hardness measurement was carried out for HP Sr-122 tapes to investigate the change of core density and the result is shown in Fig. 4(d). The $Hv$ value for HP850 tape samples is around 156.2, which is much larger than that of the flat-rolled Sr-122 tapes[11]. The high core density ensures large electrical pathway chain in polycrystalline bulk. However, the $Hv$ values for all hot-pressed tapes are almost the same around 154.0 in spite of HP temperature, suggesting that the core density seems to be saturated within the experimental conditions. Thus, it can be implied that the hot pressing can densify the Sr-122 phase, but the core density is not further enhanced by increasing HP temperature. In this work, the HP temperature has obvious effect on c-axis texture and resistivity, but little on the core density. Lastly, we show the HP-temperature dependence of the transport $J_c$ at 10 T and 4.2 K in Fig. 4(a). The variation tendency of $J_c$ values is qualitatively similar to those of $F$ and $RRR$ values. Compared to HP850 tapes, we may conclude that the $J_c$ increase for HP900 tapes is mainly attributed to higher degree of c-axis texture and enhanced grain connectivity.

In order to further figure out the reasons for the high $J_c$ values for our HP Sr-122 tapes, we performed micro-structural analysis of the Sr-122 phase by SEM characterization, as shown in Figs. 5(a-d). There are very few pores in polycrystalline core areas, which are in consist with the high $Hv$ values. In Fig. 5(a), in some cases, we can see some irregular morphology and isolated particles of Sr-122 phase for HP850 samples. In Figs. 5(b-c), when the HP temperature increases to 875 and 900 $^o$C, the planar grains become relatively more flat and most of grains seem to be well connected. This indicates good grain alignment and coupling, and is in accordance with the higher $RRR$ and $F$ values. Fig. 5(d) exhibits typical SEM image of the samples hot-pressed at 925 $^o$C. A further melted and smooth planar structure can be observed, a common phenomenon for bulks processed at high temperature. However, microstructure reveals that some small black regions and many micro-cracks (marked by red arrows) emerged in the core area, which may counteract the positive effects on c-axis texture and grain coupling. The HP temperature of 925 $^o$C is very close to the melting point (961 $^o$C) of Ag, and thus the softness of Ag sheath is easily induced. The external pressure during heat treatment can cause severe deformation of the superconducting core, which leads to micro-cracks. Because of such more micro-cracks, the transport $J_c$ of HP925 samples is greatly reduced.

A useful tool to clarify the grain size, local orientation of the grains and misorientation angles between grains in high-$T_c$ superconductors is EBSD technique, which enables the orientation mapping of granular samples by means of automatized recording of Kikuchi patterns[24-26]. We performed the EBSD scans on the polished center sections of superconducting cores for both HP850 and HP900 tapes. As shown in Figs. 6(a) and (c), the inverse pole figure (IPF) images presents the crystallographic orientation in [0 0 1] direction found in polished core areas. The dominant orientation is (001) as the expected red color for both tapes, but there is a small (100) orientation for HP850 tapes as the green color. The crystal-direction analysis reveals that about 9.1% and 25.7% of Sr-122 grains is oriented within the limit of $10^o$ around [001] direction in HP850 and HP900 tapes, respectively. And about 41.8% and 26.2% of Sr-122 grains is oriented within the range of $10^o$-$20^o$ in HP850 and HP900 tapes, respectively. The result is also confirmed by pole figures (See Supplemental Figs. S1 and S2) that our HP Sr-122 tapes are much textured materials and HP900 samples have higher degree of c-axis grain alignment. Engineering such textured materials is an effective method to overcome the weak-linked problem and improve intergrowths between grains[9, 27]. However, the grain size variation is huge since the grain size is as large as 4-7 μm in areas A and C or smaller than 1 μm in areas B and D. Especially for HP850 tapes, many small grains (<500 nm) are presented in misorientation angles image (see Fig.6 (b)), which is in agreement with the R-T and SEM observations confirming relatively inferior connection between grains compared to HP900 samples. On the other hand, fully oriented grains do practically not exist in polycrystalline material, which implies that a current flowing through the tape has to face a large amount of obstacles formed by high-angle GBs[28, 29]. Figs.6 (b) and (d) show boundary misorientations between $2^o$-$5^o$ (green), $5^o$-$10^o$ (blue), $10^o$-$30^o$ (yellow) and $30^o$-$90^o$ (red). Many small-angle boundaries (misorientations smaller than $5^o$ shown in green) are detected for both samples. It is found that the fraction of misorientation angles between 2-$10^o$ for HP900 tapes (~26.2%) is larger than that of HP850 tapes (~23.3%). The large fraction of small misorientation angles is an important factor to improve superconducting properties of HP900 samples.

To clarify how the high transport $J_c$ is achieved in our pressed Sr-122 tapes, the center section of superconducting core was submitted to transmission electron microscopy (TEM) examination. As exhibited in Fig. 7(a), the TEM observation confirms clean and well-connected GBs in a polycrystalline bulk. The high-resolution TEM (HRTEM) images reveal many small misorientaion angles, as shown by a typical GB in Fig. 7(b). Previous results demonstrated that the elements of superconducting phase homogeneously distributed throughout the superconducting core of Sr-122 tapes[11, 12]. Here, we further studied the changes of chemical composition between various grains. Fig. 7(c) shows a typical image of four grains (marked as C, D, E and F) in the hot-pressed tapes. The element distribution of Sr-122 phase is almost homogeneously dispersed in these four Sr-122 grains and that there is no obvious element deposition in GBs (Supplemental Figs. S3). This phenomenon can effectively reduce the local suppression of order parameter at GBs[28, 29]. More notably, the TEM examination in Fig. 7(a) reveals that there are a number of dislocations in Sr-122 phase. The HRTEM micrograph in Fig. 7(d) clearly shows the distortions of lattice. It is well known that the maximum electrical current density in high $T_c$ superconductors is ultimately determined by the vortex pinning defects that can be tailored into the material without degrading the electrical pathway[30, 31]. Therefore, these small lattice dislocations can serve as strong pinning centers to improve flux pinning force and thus enhance $J_c$-$B$ performance, similar to the result of $MgB_2$ and YBCO conductors[32, 30].

**Discussion**

The 122-type iron-pnictides have complex chemical composition and relatively hard phase. The fabricating schedules were previously developed by several groups for making high quality Ba-122 or Sr-122 samples[8-13, 33]. It is clear that the final sintering temperature definitely plays an important role on the $J_c$–$B$ properties of 122-type iron-pnictides. Here, we prepared a series of HP Sr-122 tapes at different HP temperatures (850-925 $^o$C), and high $T_c$, narrow transition width and high $Hv$ value manifest the good quality of superconducting phase. The best HP900 tapes have an excellent transport $J_c$-$B$ performance. It is thought that high degree of c-axis texture and improved grain connectivity are responsible for the significant $J_c$ enhancement.

Firstly, the HP900 tapes have the extremely high texture parameter $F$ of ~0.58. The value is much larger than those of HP850 and HP875 tapes. This value is also higher than those of previous hot-pressed Sr-122 and cold-pressed Ba-122 tapes[11-13]. The strong c-axis texture guarantees the large proportion of the low-angle GBs. It is reported that the transport $J_c$ will not suffer much depression when it run across GBs with small misoriention angles < 9°C in Ba-122 bicrystal[34]. As a result, a superior inter-grain $J_c$ can be obtained for HP900 tapes. Secondly, the electrical resistivity-temperature curves for HP samples display a high ratio of room-temperature and normal-sate resistivity[16, 17]. The residual resistivity ratio $RRR$ values for all tapes are well over 4.2, which are much larger than the flat-rolled Sr-122/Fe tapes and typical MgB$_2$ conductors[22, 19, 20]. The HP900 samples have lowest residual resistivity $\rho(40K)$ and highest residual resistivity ratio $RRR$, indicating good re-crystallization. This means that careful control of the HP temperature can bring a complete reaction, resulting in better grain connection and ultimately higher $J_c$[9, 35]. EBSD images further confirm that the small grains (<500 nm) are greatly reduced in HP900 samples compared to HP850 tapes. Therefore, the good grain connection is considered to be another origin of the $J_c$ enhancement. Consequently, large $J_c$ up to ~$10^5$ A/cm$^2$ in a high field of 14 T and at 4.2K has been achieved in our Sr-122 tapes. The value is superior to those of typical NbTi and MgB$_2$ conductors, which have already realized commercial applications. This advanced $J_c$ result clearly strengthens the position of iron-based conductors as a competitor to the conventional and MgB$_2$ superconductors for high field applications.

On the other hand, the reason of the high transport $J_c$ in our hot-pressed tapes is also discussed by TEM observation. Both HP850 and HP900 tapes show clean GBs and high density of dislocations in Sr-122 phase (See Fig. 7 and Supplemental Fig. S4). The flux pinning and the behavior of GBs for HP850 samples are similar to those of HP900 tapes, indicating that there are common phenomena for our hot-pressed Sr-122 tapes. This may be another important reason for high $J_c$-$B$ performance. However, inhomogeneous distribution of grain size is still existed in present samples. Further improvement in $J_c$ can be expected upon optimizing process parameters, such as reducing pressing time and employing slow cooling rate.

**Methods**

**Sample preparation.** We fabricated Ag-clad $Sr_{0.6}K_{0.4}Fe_2As_2$ tapes using Sn as additive by the *ex-situ* PIT method. Sr fillings, K pieces, and Fe and As powder with a ratio of Sr:K:Fe:As = 0.6: 0.5: 2: 2.05 were mixed for 12 hours by ball-milling method. The milled powders were packed into Nb tubes and then sintered at 900 °C for 35 h. As prepared Sr-122 superconducting precursors were then ground into powders under Ar atmosphere. In order to increase grain connectivity, the precursors were mixed with 5 wt% Sn by hand with an agate mortar. Then the fine powders were packed into Ag tubes with OD 8 mm and ID 5 mm. These tubes were sealed and then cold worked into tapes (~ 0.4 mm thickness) by swaging and flat rolling. Finally, hot pressing was performed on the 60 mm long tapes under ~30 MPa at the sintering temperature 850, 875, 900 and 925 °C for 30 min. These tapes are defined as HP850, HP875, HP900 and HP925 tapes, respectively. Further experiment details were described elsewhere[11, 12].

**Measurements.** Phase identification of samples was characterized by X-ray diffraction (XRD) analysis with Cu Kα radiation. Resistivity measurements of the superconducting core were carried out using a PPMS system. Microstructure characterization was analyzed using SEM, EBSD and TEM images. Vickers hardness of the tape samples was measured on the polished cross sections with 0.05 kg load and 10 s duration in a row at the center of the superconducting cross section. The transport critical current $I_c$ was measured at 4.2 K using short tape samples of 3 cm in length with the standard four-probe method and evaluated by the criterion of 1 μV/cm, then the critical current was divided by the cross section area of the superconducting core to get the critical current density $J_c$. The applied fields up to 14 T in transport $I_c$ measurement were parallel to the tape surface.

**Acknowledgments**

This work is partially supported by the National '973' Program (grant No. 2011CBA00105) and the National Natural Science Foundation of China (grant Nos. 51025726, 51320105015 and 51202243).

**Captions**

Figure 1 Bulk XRD patterns for the superconducting cores of the hot-pressed Sr-122 tapes prepared at 850, 875, 900 and 925 $^{o}$C. As a reference, the data for randomly orientated powder is also included. The peaks of $Sr_{0.6}K_{0.4}Fe_2As_2$ phase are indexed, while the peaks of Ag and AgSrAs phases are also marked. The Ag peaks are contributed from the Ag sheath.

Figure 2 The resistivity versus temperature curves of the hot-pressed Sr-122 tapes prepared at 850, 875, 900 and 925 $^{o}$C. All data were obtained after peeling off the Ag sheath.

Figure 3 Magnetic field dependence of transport $J_c$ at 4.2 K for HP Sr-122 tapes with different HP temperatures. The applied fields up to 14 T were parallel to the tape surface.

Figure 4 The HP-temperature dependence of the transport $J_c$ (a), c-axis texture parameter $F$ (b), residual resistivity ratio $RRR$ (c) and average Vickers micro-hardness $Hv$ values (d). The $F$(T), $RRR$(T) and $J_c$(T) values are obtained from the above Figs. 1, 2 and 3, respectively.

Figure 5 SEM images of the superconducting cores: including the Sr-122 samples hot-pressed at 850 $^{o}$C (a), 875 $^{o}$C (b), 900 $^{o}$C (c) and 925 $^{o}$C (d). The micro-cracks in (d) are marked by red arrows.

Figure 6 EBSD images for HP850 and HP900 samples: (a) inverse pole figure (IPF) image in [001] direction of HP850 tapes, (b) boundary misorientation of HP850 tapes, (c) IPF image in [001] direction of HP900 tapes and (d) boundary misorientation of HP900 tapes. The EBSD scans were performed on the polished center sections of superconducting cores in a/b plane. The color codes in (a) and (c) are explained in the right stereographic triangles. The different degree of boundary misorientations in (b) and (d) are indicated by green line (2$^{o}$-5$^{o}$), blue line (5$^{o}$-10$^{o}$), yellow line (10$^{o}$-30$^{o}$) and red line (30$^{o}$-90$^{o}$).

Figure 7 TEM study on the GBs and defects of HP900 tapes. (a) TEM observation showing clean GBs and a high density of dislocations (marked as red arrows)

in the polycrystalline bulk. (b) HRTEM image showing a small misorientaion angle (~ 4 deg.) of two grains (marked as A and B). (c) The detailed element distribution of superconducting phase (four grains marked as C, D, E and F) is analyzed by EDS mapping. (d) HRTEM image showing lattice distortions in Sr-122 grains. The TEM scans were performed on the center sections of superconducting cores processed by focused ion beam (FIB) technique.

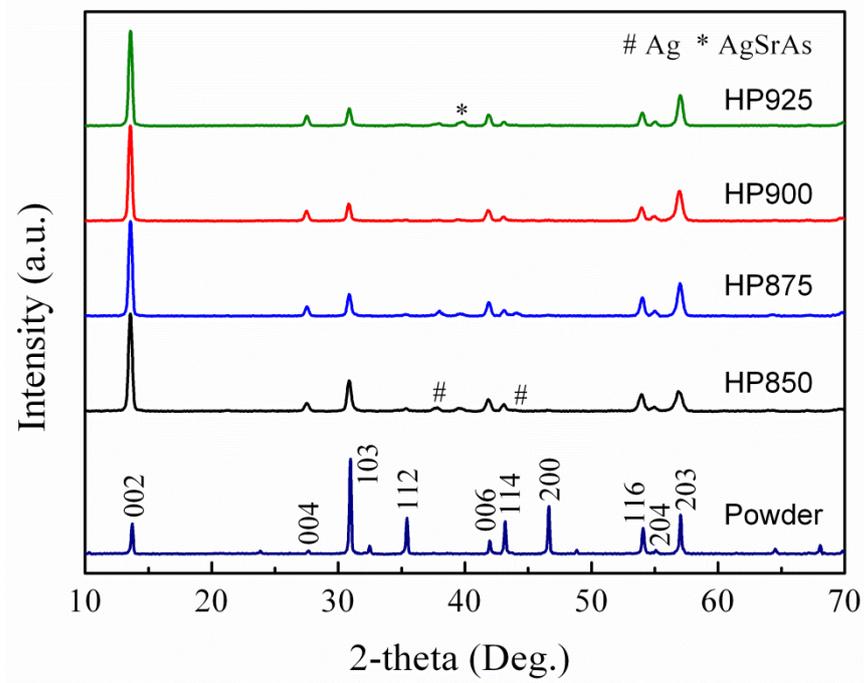

Figure 1 Lin et al.

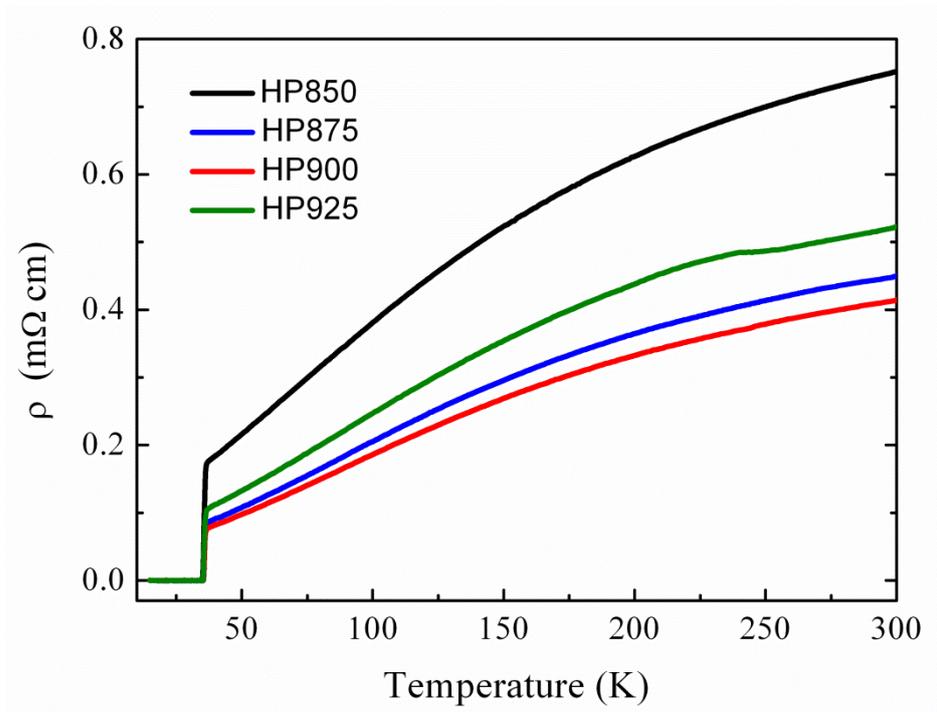

Figure 2 Lin et al.

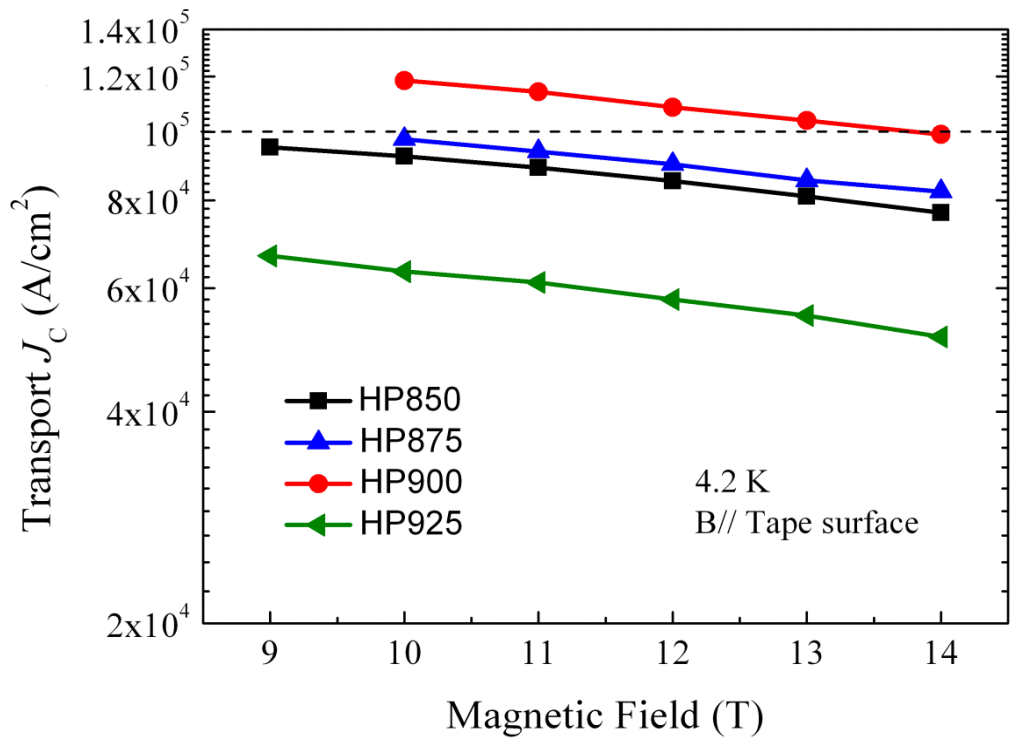

Figure 3 Lin et al.

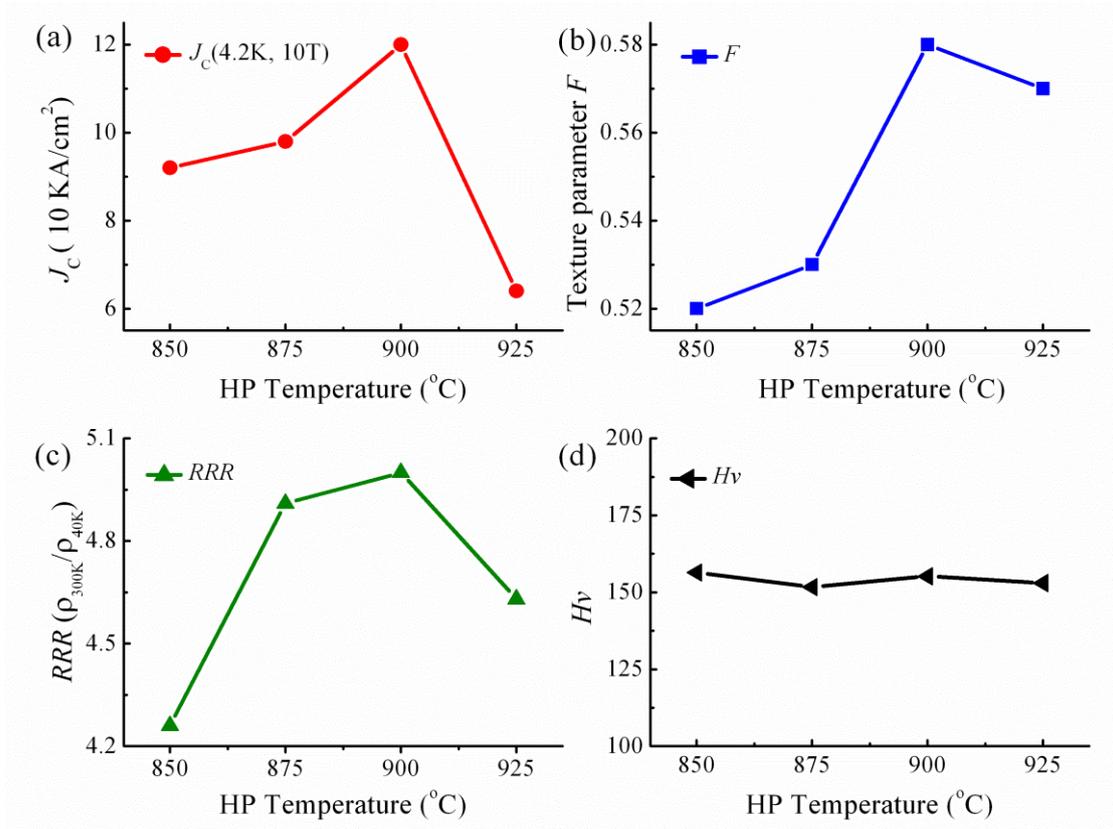

Figure 4 Lin et al.

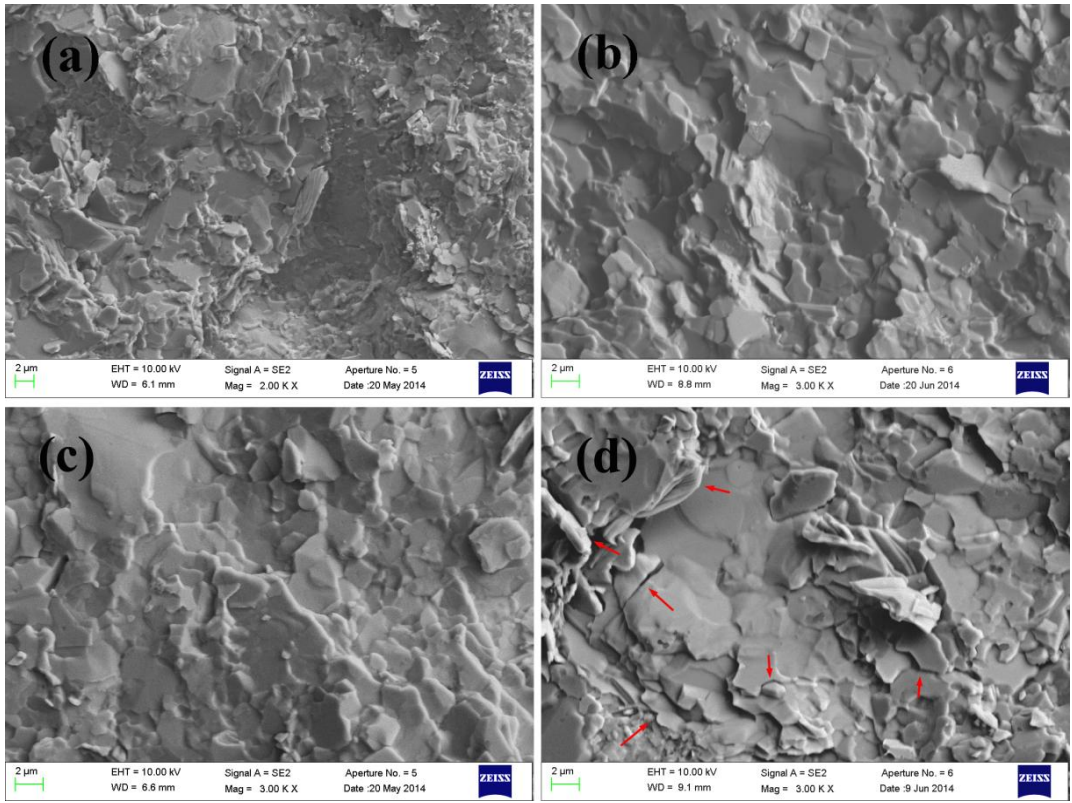

Figure 5 Lin et al.

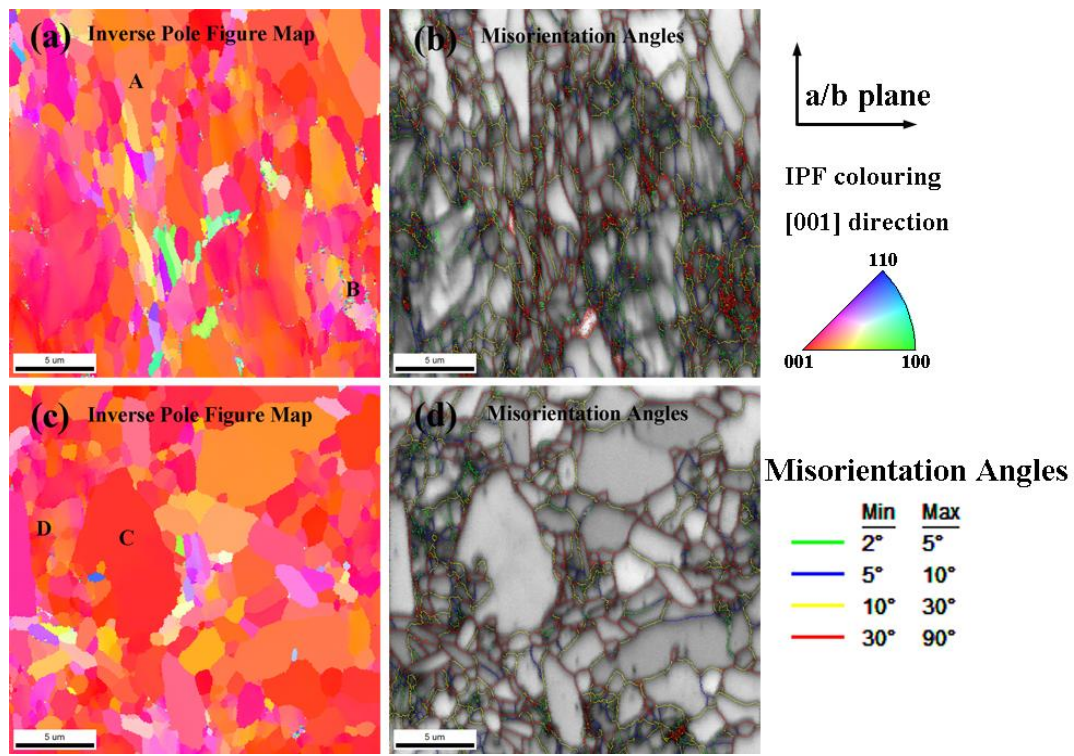

Figure 6 Lin et al.

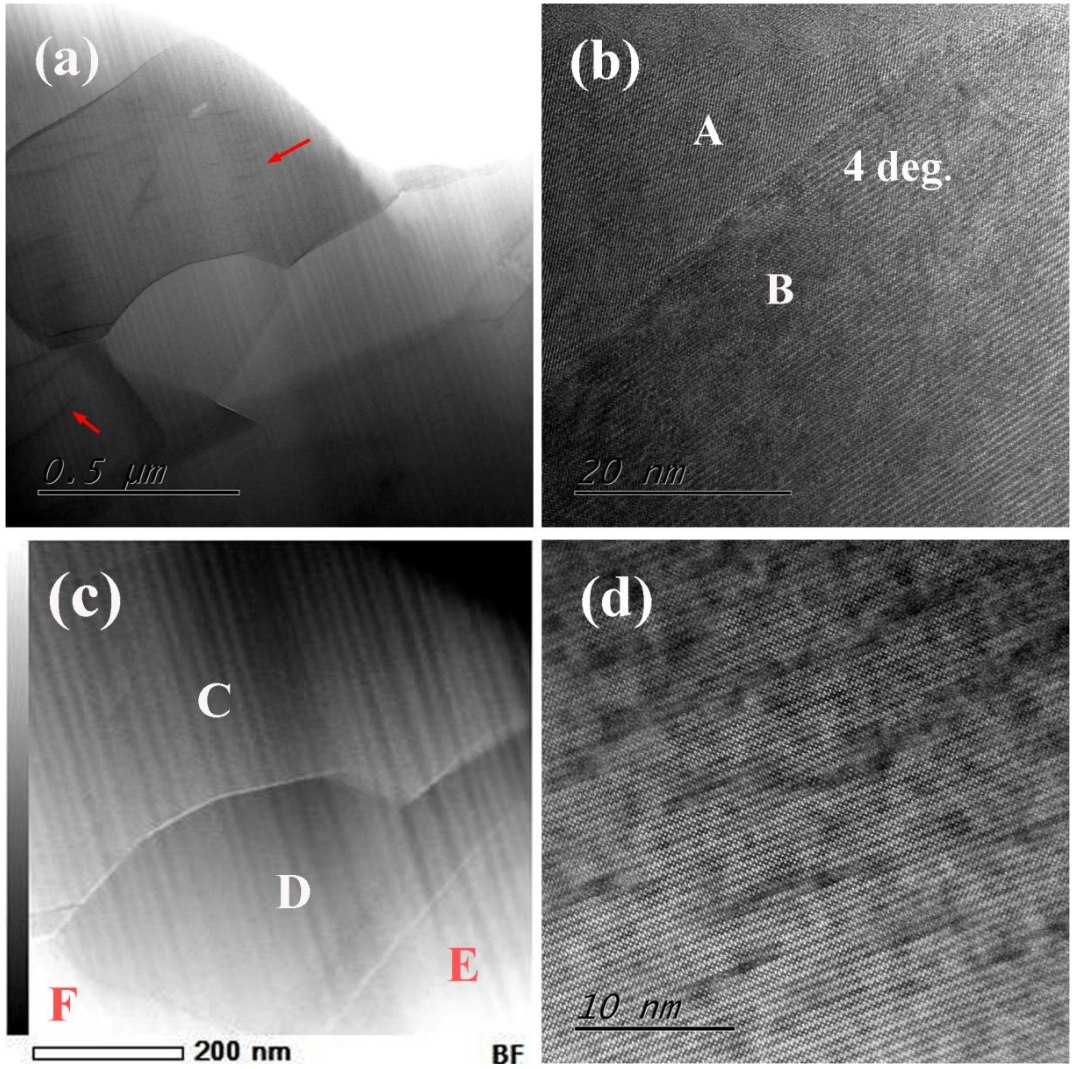

Figure 7 Lin et al.